\documentstyle[12pt]{article}

\begin{document}

\begin{center}

{\Large \bf The Velocity of Gravitational Waves} \\

\vspace{1.8cm}

M. Novello$\mbox{}^{*}$, V. A. De Lorenci, Luciane R. 
de Freitas$^{\dag}$ and O. D. Aguiar$^{\ddag}$\\
\vspace{0.8cm}
{\it Centro Brasileiro de Pesquisas F\'{\i}sicas,} \\
\vspace{0.1cm}
{\it Rua Dr. Xavier Sigaud, 150, Urca} \\
\vspace{0.1cm}
{\it 22290-180 -- Rio de Janeiro, RJ -- Brazil.} \\
\vspace{0.8cm}
$\dag${\it Instituto de F\'{\i}sica,} \\
\vspace{0.1cm}
{\it Universidade Federal do Rio de Janeiro} \\
\vspace{0.1cm}
{\it Ilha do Fund\~ao - CT - Bloco A,} \\
\vspace{0.1cm}
{\it 21945 -- Rio de Janeiro, RJ --  Brazil.} \\
\vspace{0.8cm}
$\ddag${\it Instituto Nacional de Pesquisas Espaciais,} \\
\vspace{0.1cm}
{\it Divis\~ao de Astrof\'{\i}sica} \\
\vspace{0.1cm}
{\it Av. Astronautas, 1758} \\
\vspace{0.1cm}
{\it 12227-010 -- S\~ao Jos\'e dos Campos, SP -- Brazil.}
\vspace{2.0cm}

\begin{abstract}
We examine the propagation of gravitational waves 
in new field theory of gravitation 
recently proposed by Novello-De Lorenci-Luciane(NDL). This examination is 
done on a solvable case
corresponding to a spherically symmetric static configuration. 
We show that in NDL theory the velocity of gravitational waves is lower 
than light velocity. 
We point out some consequences of this result and suggest a possible 
scenario for its verification.
\end{abstract}

\end{center}

\vspace{1cm}

PACS numbers: 04.50.+h; 04.30.Nk; 04.90.+e.

\vfill{\footnotesize \tt \mbox{}\\
\line(1,0){180}
\par $\mbox{}^{*}$ novello@lafex.cbpf.br}
\newpage

\section{Introduction}
\label{introduction}

The equivalence principle (EP), which states that all kinds of matter 
(including massless particles such as photons) interact in 
the same way with 
the gravitational field, gave to Einstein the possibility to treat 
gravitational phenomena as a modification of the spacetime geometry. 

Recently we have exploited some consequences of a field theoretical
description of gravity with a new ingredient: 
we do not require that the gravitational energy satisfies the hypothesis
of universality of the equivalence principle.
This means that gravity does not couple to itself in the same way 
as all others forms of
energy. The hypothesis that gravity-gravity interaction is the same
as the interaction that 
gravity has with all other type of matter, as  made in General Relativity, 
is not supported by direct observation, but by indirect consequences of
the effect of gravitational field on matter.

The NDL theory of gravity contains many 
of the ingredients of GR and is indistinguishable
as far as standard post-Newtonian tests and 
quadrupole emission are concerned (see ref. \cite{NDL}). The key point 
to distinguish the NDL theory from GR seems to be the 
velocity of gravitational waves. 
In order to examine the behavior of such waves we will study the evolution 
of discontinuities of the corresponding equations of motion through a 
characteristic surface $\Sigma$, using the Hadamard method.

There is a general expectation concerning the possibility that we could 
detect gravitational waves in the next decade. Such situation is based on 
the great number of new experimental devices that 
many laboratories, throughout the world, are constructing. Many scientists 
are going to become involved in this enterprise.
In this vein, there is a natural and direct way to test NDL theory by the 
analysis of the gravitational waves. We 
start to undertake this task in this present paper.

In the section \ref{ShortNDL} we review the NDL theory showing its formal 
structure. We show that the gravitational radiation propagates 
on a effective metric distinct from the universal metric seen by the rest 
of matter and the static spherically symmetric solution.

In the section \ref{wave} we make a 
brief review of Hadamard's method to study wave propagation. 

In the section \ref{4} we compare the velocity of 
electromagnetic waves in the vacuum and Gravitational waves 
in the NDL theory on the static 
spherically symmetric metric and show that in this theory the velocity of 
gravitational radiation is less than c [We use units of light 
velocity ($c=1$)].

\section{A Short Review of the NDL Theory of Gravity}
\label{ShortNDL}

\subsection{General Features}

In a previous paper \cite{NDL} we presented a modification of the 
standard Feynman-Deser approach of field theoretical derivation of 
Einstein\rq s general relativity, which led to a competitive 
gravitational theory. 
The main lines of NDL approach can be 
summarized as follows:
\begin{itemize}
 \item{Gravity is described by a symmetric second rank tensor 
$\varphi_{\mu\nu}$ that satisfies a non-linear equation of motion;}
 \item{Matter couples to gravity in an universal way. In this interaction, 
the gravitational field appears only in the combination 
$ \gamma_{\mu\nu} + \varphi_{\mu\nu}$, inducing us to define a quantity
$ g_{\mu\nu} = \gamma_{\mu\nu} + \varphi_{\mu\nu}$. This tensor 
$g_{\mu\nu}$ acts as an effective metric tensor of the 
spacetime as seen by matter 
or energy of any form except gravitational energy;}
 \item{The self interaction of the gravitational field break the 
universal modification of the spacetime geometry.}
\end{itemize}

We define a three-index tensor $F_{\alpha\beta\mu}$, which we will call 
the {\bf gravitational field}, in terms of the symmetric standard variable 
$\varphi_{\mu\nu}$ (which will be treated as the potential) to describe a
spin-two field, by the expression\footnote{We are using the 
anti-symmetrization symbol $[x, y] \equiv xy - yx$ and the symmetrization symbol  $(x, y) \equiv xy + yx$. Note that indices are raised and lowered by the background metric $\gamma_{\mu\nu}$. The covariant
derivative is denoted by a semicomma `$;$' and it is constructed 
with this metric.}
\begin{equation}
F_{\alpha\beta\mu} = \frac{1}{2} ( \varphi_{\mu[\alpha;\beta]} + 
F_{[\alpha}\gamma_{\beta]\mu} )
\label{d1}
\end{equation}
where $F_{\alpha}$ is the trace 
\begin{equation} 
F_{\alpha} = F_{\alpha\mu\nu} \gamma^{\mu\nu} = \varphi_{,\alpha} - 
\varphi_{\alpha\mu;\nu} \gamma^{\mu\nu}.
\end{equation}  
From the above definition it follows that $F_{\alpha\beta\mu}$ is 
anti-symmetric in the first pair of indices and obeys the cyclic 
identity, that is
\begin{equation}
F_{\alpha\mu\nu} + F_{\mu\alpha\nu} = 0
\label{d2}
\end{equation}
\begin{equation}
F_{\alpha\mu\nu} + F_{\mu\nu\alpha} + F_{\nu\alpha\mu} = 0.
\label{d3}
\end{equation}

In order to construct a non-linear theory for the gravitational field 
we make the hypothesis that gravity-gravity interaction is represented 
by a functional of the two invariants of the field $F_{\alpha\beta\mu}$, 
A and B, given by:
\begin{eqnarray} 
A &\equiv& F_{\alpha\mu\nu}\hspace{0.5mm} F^{\alpha\mu\nu},
\nonumber \\
B &\equiv& F_{\mu}\hspace{0.5mm} F^{\mu}. 
\label{AB}
\end{eqnarray}
and the Lagrangian density is a Born-Infeld type
\begin{equation} 
L = \frac{b^2}{\kappa} \left\{ \sqrt{1 - \frac{U}{b^{2}}} 
 - 1 \right\}, 
\label{Lag}
\end{equation}
where $b$ is a constant and $U$ is defined by 
\begin{equation}
U\equiv A-B.
\label{U}
\end{equation}
In this way, the gravitational action is expressed as:
\begin{equation}
S = \int{\rm d}^{4}x\sqrt{-\gamma}\,L,
\label{action}
\end{equation}
where $\gamma$ is the determinant of the flat spacetime 
metric $\gamma_{\mu\nu}$ written in an arbitrary coordinate system.

From the Hamilton principle we find the following equation of motion 
in the absence of material sources:
\begin{equation}
\left[L_{U} F^{\lambda (\mu\nu)} \right]_{;\lambda} = 0.
\label{eqmov}
\end{equation}
$L_{U}$ represents the derivative of the Lagrangian 
with respect to the invariant $U$. 

As we remarked before, we follow the standard 
procedure \cite{Feynman,Deser} and define an 
effective Riemannian metric tensor in terms of the potential 
$\varphi_{\alpha\beta}$, by the expression
\begin{equation}
g_{\mu\nu} = \gamma_{\mu\nu} + \varphi_{\mu\nu}.
\label{b1}
\end{equation}
This relation has a deep meaning, for all forms of non-gravitational 
energy the net effect of the gravitational field is felt precisely as if 
gravity were nothing but a consequence 
of changing the metrical properties of the spacetime 
from the flatness structure to a curved one. The definition of the
associated metric tensor is provided 
precisely by the above expression. This means that any 
material body (including those massless) moves along geodesics as if 
the metric tensor of spacetime were given by the equation (\ref{b1}). 
This is in agreement with the Equivalence Principle.

\subsection{The Static Spherically Symmetric Solution}

We set for the auxiliary metric\footnote{We note that this metric
is non-observable by any form of energy.} of the background the form
\begin{equation}
{\rm d}s^{2} = {\rm d}t^{2} - {\rm d}r^{2} - r^{2} ( {\rm d}\theta^{2} 
+ \sin^{2}\theta {\rm d}\varphi^{2} ).
\protect\label{ds1}
\end{equation}
This means that all operations of raising and lowering indices are made by 
this metric $\gamma_{\mu \nu}$. We remark the fact that 
matter\footnote{Massive or massless particles --- photons, for instance --- 
that is, any form of non-gravitational energy} feels a modified geometry 
given by Eq.
(\ref{b1}).

A  possible static spherically symmetric solution of 
NDL theory has only two non zero gravitational components of
$\varphi^{\mu\nu}$:
\begin{eqnarray}
\varphi_{00} &=&  \varphi^{00} = \mu(r)
\label{phi00}  \\
\varphi_{11} &=&   \varphi^{11}  = - \nu(r). 
\label{phi11}
\end{eqnarray} 
In this case the gravitational field $F_{\alpha\beta\mu}$ reduces to
\begin{eqnarray}
F_{100} &=& -  \frac{\nu}{r}\\
F_{122} &=& \frac{F_{133}}{\sin^{2}\theta} =  \frac{1}{2} \nu r - \frac{1}{2} \mu' r^2  \\
\end{eqnarray}
in which a prime $'$ symbolizes the derivative with respect 
to the radial variable $r$. The unique component of the trace that 
remains is $F_{1}$,
\begin{equation}
F_{1} =  \mu^{'} - 2\hspace{0.5mm} \frac{\nu}{r}.
\end{equation}

From these we can evaluate
 the invariant $U$:
\begin{equation}
U = \frac{\nu^2}{r^2} - \frac{2\nu\mu'}{r}.
\label{Usol}
\end{equation}

From the equations of motion (\ref{eqmov}) we obtain 
\begin{eqnarray}
\nu = \hspace{0.9mm} \frac{2M}{r}\hspace{0.9mm} 
\left\{ 1 - {(\frac{r_{c}}{r}})^{4} \right\}^{-\frac{1}{2}}
\label{nusol}\\
\mu = \frac{1}{2} \sqrt{b M} \left\{ F(\alpha, \sqrt{2}/2 ) + 
\mu_{0} \right\}
\label{musol}
\end{eqnarray}
in which the constant $r_{c}$ appearing in Eq. (\ref{nusol}) is given by
\begin{equation}
{r_{c}}^{2}  \equiv \frac{2M}{b}. 
\end{equation}
In Eq. (\ref{musol}) $F(\alpha, \sqrt{2}/2 )$ is the elliptic 
function of the first kind and the constant $\mu_{0}$ must be chosen 
to yield the correct 
asymptotic limit. The quantity $\alpha$ is given by 
\begin{equation}
\alpha \equiv \arcsin \left[ 1 - \left(\frac{r_c}{r}\right)^{2} 
\right]^{\frac{1}{2}} .
\end{equation} 

\section{Wave Propagation}
\label{wave}
In this section we discuss wave propagation in two 
steps. First we give a very intuitive example using the 
electromagnetic field. We make a brief review of Hadamard's 
method and show why in non-linear cases it is better apply this 
treatment. Second we use this method on the NDL theory of gravitation

\subsection{Hadamard Treatment}
\label{hadamard}

Let us consider a charged particle at rest for $t \leq t_{0}$ that starts to be accelerated in $t = t_{0}$. The electromagnetic field produced by this particle was initially independent of time and after that it becomes a variable field. Hence, there are two regions of space-time one in which the field is independent of the time and one in which the field  varies. These regions are separated by a hypersurface  
$\Sigma$. As the accelerated particle emits radiation, the hypersurface $\Sigma$ is the corresponding wavefront.

Mathematically such a hypersurface $\Sigma$ is characterized by the fact that some derivatives of the electromagnetic potential $A_{\mu}$ are discontinuous 
across $\Sigma$. This effect comes from the fact that all derivatives of the 
potential $A_{\mu}$ are zero in the region $\Sigma^{-}$ but not in $\Sigma^{+}$. These disturbances are propagated by the equation of motion.

Although the situation described above is a good image for any field  disturbance, the equation of motion of the gravitational field is non-linear and this fact causes a great difference between the analysis of the propagation of a gravitational wave and of an electromagnetic wave. Due to this, the better and simpler method to study the propagation of the gravitational wave is Hadamard's method, which we explain briefly in the following.

Let us define a hypersurface of discontinuity $\Sigma$ by the  equation:
\begin{equation}
\Sigma : \varphi(x^{\alpha}) = 0.
\label{phi}
\end{equation}
The space-time is divided in two regions: $\Sigma^{+}$ (where $\varphi > 0$) and $\Sigma^{-}$ (where $\varphi < 0$). The discontinuity of any function $f(x^{\alpha})$ across $\Sigma$ is defined by:
\begin{equation}
[f(x^{\alpha})] = \lim_{p^{\pm}\rightarrow p} \left( f(P^{+}) - f(P^{-}) \right)
\label{defdes}\end{equation}
where $P^{+}$ is in $\Sigma^{+}$ and $P^{-}$ is in $\Sigma^{-}$.
To introduce the algebraic formalism proposed by
Hadamard we use the 
definition of the total differential of a function f, to state that

\begin{equation}
[df] = 0
\label{descdadif}\end{equation}
since
$$ [df] = \lim_{p^{\pm} \rightarrow p} \left(df(P^{+}) - df(P^{-})\right) $$
It follows

\begin{equation}
[df] = [\frac{\partial f}{\partial x^{\alpha}} dx^{\alpha}] = 0 \label{demos2}
\end{equation}
but $ x^{\alpha} $ 
is a coordinate and since the coordinate system is continuous, $ dx^{\alpha}$ is also, hence,
\begin{equation} 
 [\frac{\partial f}{\partial x^{\alpha}}] dx^{\alpha} = 0
\label{demos3}\end{equation}
this differential is calculated on the surface therefore $dx^{\alpha}$ is parallel to it 
\begin{equation}
[\frac{\partial f}{\partial x^{\alpha}}] \propto \varphi_{,\alpha} \equiv K_{\alpha}
\label{demos4}\end{equation}
where $\varphi_{,\alpha}$ is normal to hypersurface $\Sigma$. This result is valid for tensorial functions of any rank. We can verify this 
using an suitable transformation of coordinates for a  particular case of a vector and it is straightforward extended to higher rank tensor.

\subsection{Wave Propagation in NDL Theory}
\label{wave ndl} 
Let us review briefly the result set up in \cite{NDL} for the 
velocity of the gravitational wave in NDL theory. We represent 
by the symbol $[ J ]_{\Sigma}$ the 
discontinuity of the function $J$ across the surface $\Sigma$.
Following Hadamard we impose the discontinuity conditions 
for the field:
\begin{equation}
[F_{\mu\nu\alpha}]_{\Sigma} = 0
\label{gw6}
\end{equation} 
and   
\begin{equation}
[F_{\mu\nu\alpha;\lambda}]_{\Sigma} = f_{\mu\nu\alpha} k_{\lambda},
\label{gw7}
\end{equation}
where $k_{\alpha}$ represents the normal vector to the surface of 
discontinuity $\Sigma$. $f_{\alpha\beta\gamma}$ has the same 
symmetries as the field $F_{\alpha\beta\gamma}$.
Taking the discontinuity of the equation of motion (\ref{eqmov}) 
we obtain:
\begin{equation}
f^{\mu}\mbox{}_{(\alpha\beta)} k_{\mu} + 2 \frac{L_{UU}}
{L_{U}} ( \eta - \zeta ) F^{\mu}\mbox{}_{(\alpha\beta)} k_{\mu} = 0
\label{gw8}
\end{equation}
in which we defined $\eta \equiv F_{\alpha\beta\mu} f^{\alpha\beta\mu}$ and
$\zeta \equiv F_{\mu} f^{\mu}$.
In the same way, taking the discontinuity of the indentity
\begin{equation}
F_{\alpha\beta}\mbox{}^{\lambda}\mbox{}_{;\gamma} +
F_{\beta\gamma}\mbox{}^{\lambda}\mbox{}_{;\alpha} +
F_{\gamma\alpha}\mbox{}^{\lambda}\mbox{}_{;\beta} =
\frac{1}{2}\delta^{\lambda}_{[\alpha}F_{\gamma ];\beta} + 
\frac{1}{2}\delta^{\lambda}_{[\gamma}F_{\beta ];\alpha} +
\frac{1}{2}\delta^{\lambda}_{[\beta}F_{\alpha ];\gamma}
\end{equation}
contracted with $F^{\alpha\beta\lambda}k^{\gamma}$, we obtain: 
\begin{equation}
(\eta-\zeta)k^2 + 2F^{\alpha\beta\lambda}f_{\beta\gamma\lambda}
k_{\alpha}k^{\gamma} + F^{\alpha\beta\lambda}f_{\alpha}
k_{\beta}k^{\lambda} + F^{\beta}f^{\gamma}
k_{\beta}k^{\gamma} = 0.
\label{eqtres}
\end{equation}
Working with equations (\ref{gw8}) and (\ref{eqtres}), we arise at the
following expression for the propagation: 
\begin{equation}
k_{\mu} k_{\nu} [ \gamma^{\mu\nu} + \Lambda^{\mu\nu} ] = 0
\label{51}
\end{equation}
in which the quantity  $\Lambda^{\mu\nu}$ is written in terms of the 
gravitational field:
\begin{equation} 
\Lambda^{\mu\nu} \equiv 2 \frac{L_{UU}}{L_{U}} 
[ F^{\mu\alpha\beta} 
F^{\nu}\mbox{}_{(\alpha\beta)} - F^{\mu} F^{\nu}  ]. 
\label{Lambda}
\end{equation}
Note that the discontinuities of the gravitational fields propagate in a 
modified geometry, changing the background geometry $\gamma^{\mu\nu}$ 
into an effective 
one $g^{\mu\nu}_{\scriptscriptstyle {\rm eff}}$,
\begin{equation}
g^{\mu\nu}_{\scriptscriptstyle {\rm eff}} \equiv \gamma^{\mu\nu} + \Lambda^{\mu\nu}
\label{gtilde}
\end{equation}
which depends on the field
$F_{\alpha\beta\mu}$ and also on the dynamics. This fact shows that such a 
property stems from the structural form of the Lagrangian.
Thus, in the NDL theory the characteristic surfaces of the 
gravitational waves propagate on the null cone of
an effective geometry. We remark that this geometry is  distinct from 
that observed by all other forms 
of energy and matter, which differs from the general relativity
predictions. This result gives a possibility to
choose between these two theories just by observations of the 
gravitational waves. In the next section we will present a summary
of the main properties of the solutions of the gravitational field
in NDL theory. Then we will evaluate the velocity of gravitational
wave in such background and compare with the GR result.

\section{Electromagnetic  and Gravitational Waves in the Spherically 
Symmetric Solution}
\label{4}

From what we have seen, in the NDL theory all kinds of matter and 
non-gravitational energy couple to gravity differently from gravity to 
gravity interaction. In order to compare the propagation of electromagnetic 
and gravitational waves we will proceed as follows. 

The metric that defines the structure of the spacetime
in which the discontinuities of the electromagnetic field propagate, 
is provided by Eq. (\ref{b1}). We define a 4-vector $l_{\mu}$ such that
\begin{equation}
l_{\mu}l_{\nu} g^{\mu\nu} = 0
\label{elec}
\end{equation}
where $g^{\mu\nu}$ is the inverse of the metric $g_{\mu\nu}$, defined by
\begin{equation}
g^{\mu\rho}g_{\mu\sigma} = \delta^{\rho}\mbox{}_{\sigma}.
\label{inverseg}
\end{equation}
Analogous to the section (\ref{wave}), this 4-vector is a gradient of the
hypersurface of discontinuities.
The background gravitational field we are considering here, corresponds to
a spherically symmetric and static configuration. Thus, the relation 
(\ref{elec}) reduces in this case to
\begin{equation}
\frac{\left(l_0\right)^2}{1 + \mu} - 
\frac{\left(l_1\right)^2}{1 + \nu} - 
\frac{\left(l_2\right)^2}{r^2} - 
\frac{\left(l_3\right)^2}{r^2\sin^2\theta} = 0.
\label{54}
\end{equation}
Gravitational waves propagate
as null geodesic in an effective geometry $g_{\scriptscriptstyle {\rm eff}}^{\mu\nu}$ given by 
Eq. (\ref{gtilde}). The quantity $\Lambda^{\mu\nu}$ is defined by Eq. 
(\ref{Lambda}) and have 
only two non null components:
\begin{eqnarray}
\Lambda^{00} &=& -\frac{4M^2}{b^2r^4}
\label{Lambda00}\\ 
\Lambda^{11} &=& - \Lambda^{00}.
\label{Lambda11}
\end{eqnarray} 
Correspondingly, the effective metric is provided by
\begin{eqnarray}
g_{\scriptscriptstyle {\rm eff}}^{00} &=& 1 - \frac{4M^2}{b^2r^4} 
\label{g00}\\
g_{\scriptscriptstyle {\rm eff}}^{11} &=& -1 + \frac{4M^2}{b^2r^4}
\label{g11}\\
g_{\scriptscriptstyle {\rm eff}}^{22} &=&-\frac{1}{r^2}
\label{g22}\\
g_{\scriptscriptstyle {\rm eff}}^{33} &=& -\frac{1}{r^2\sin^2\theta}
\label{g33}
\end{eqnarray}
Inserting these results in Eq. (\ref{51}) the equation of propagation of 
gravitational waves became:
\begin{equation}
\left(1 - \frac{4M^2}{b^2r^4}\right)\left(k_0\right)^2 - 
\left(1 - \frac{4M^2}{b^2r^4}\right)\left(k_1\right)^2 - 
\frac{\left(k_2\right)^2}{r^2} - 
\frac{\left(k_3\right)^2}{r^2\sin^2\theta} = 0.
\label{46}
\end{equation}

We can summarize this situation as:
\begin{itemize}
\item
Electromagnetic waves propagate in null cone of the geometry $g_{\mu\nu}$;
\item
Gravitational waves propagate in null cone of the geometry 
$g^{\scriptscriptstyle {\rm eff}}_{\mu\nu}$.
\end{itemize}

A question then arises: {\it Which wave propagates faster?}
In order to investigate this problem we can run by two equivalent ways: one
can either
evaluate the norm of the vector $l_{\mu}$ in the geometry $g^{\scriptscriptstyle {\rm eff}}_{\mu\nu}$:
\begin{equation}
l_{\mu}l_{\nu}g_{\scriptscriptstyle {\rm eff}}^{\mu\nu}
\label{lk}
\end{equation}
or else evaluate the norm of the vector $k_{\mu}$ in the geometry $g_{\mu\nu}$:
\begin{equation}
k_{\mu}k_{\nu}g^{\mu\nu}.
\label{kl}
\end{equation}
In this vein we evaluate the quantity (\ref{lk}) and investigate the 
character of 
$l_{\mu}$ in the geometry $g_{\scriptscriptstyle {\rm eff}}^{\mu\nu}$. For the solution we are
considering here, results:
\begin{equation}
||l||_{g_{\scriptscriptstyle {\rm eff}}} = \left(1 - \frac{4M^2}{b^2r^4}\right)\left(l_0\right)^2 - 
\left(1 - \frac{4M^2}{b^2r^4}\right)\left(l_1\right)^2 - 
\frac{\left(l_2\right)^2}{r^2} - 
\frac{\left(l_3\right)^2}{r^2\sin^2\theta}
\label{48}
\end{equation}
Substituting expression (\ref{54}) in the last two terms in the right hand
side of the above equation, we obtain 
\begin{equation}
||l||_{g_{\scriptscriptstyle {\rm eff}}} =
\left(1 - \frac{4M^2}{b^2r^4} - \frac{1}{1+\mu}\right)\left(l_0\right)^2 - 
\left(1 - \frac{4M^2}{b^2r^4} - \frac{1}{1+\nu}\right)\left(l_1\right)^2.
\label{56}
\end{equation}
The solutions of the functions $\mu$ and $\nu$ are given
by (\ref{musol}) and (\ref{nusol}). Since we are interested here
just in the sign of the norm of the 4-vector $l_{\mu}$ in the geometry
determined by $g^{\scriptscriptstyle {\rm eff}}_{\mu\nu}$, 
it is enough to consider only the main terms.
Thus, expanding the field solutions as: 
\begin{eqnarray}
\frac{1}{1+\mu} &\approx&  1 + \frac{2M}{r} + O\left(r^{-2}\right) 
\label{apmu}\\
\frac{1}{1+\nu} &\approx&  1 - \frac{2M}{r} + O\left(r^{-2}\right),
\label{apnu}
\end{eqnarray}
and substituting in Eq. (\ref{56}) result in
\begin{equation}
||l||_{g_{\scriptscriptstyle {\rm eff}}} = -\frac{2M}{r}\left(l_0\right)^2 
- \frac{2M}{r}\left(l_1\right)^2.
\label{61}
\end{equation}
Hence, $||l||_{g_{\scriptscriptstyle {\rm eff}}} < 0 $.
We thus conclude that $l_{\mu}$ is a space-like vector in the geometry 
$g^{\scriptscriptstyle {\rm eff}}_{\mu\nu}$, i.e., in the NDL theory
gravitational waves travel with velocity lower than light.

\section{Conclusion}

The NDL theory prediction that gravitational waves travel slower than light
has some interesting consequences. 
The first one is the possibility to exist gravitational
Ch\^erenkov Radiation, that is the emission of gravitational radiation
when a massive particle exceeds locally the ``graviton'' speed \cite{Caves}.
Evidently, this phenomenon will put limits on how far from their
sources cosmic rays can be found with ultra high energies, such as
the ones searched by the AUGER Project (E $\ge 10^{20}$eV) \cite{Escobar}. 
How stringent these limits are will be addressed in future work.

One question naturally stands up: can we obtain observational evidence
that gravitational waves travel at speeds below the light velocity 
in the presence
of gravitational potentials? Very high gravitational potentials can
be found in the vicinities of neutron stars, black holes and
supernova cores. However, unless we are
talking about very massive black holes, all these gravitational
sources are not sufficiently extended in size to allow the integration
of a large enough difference in the arriving times at Earth for
photons and gravitons supposely generated at the same place and
instant. On the other hand, black holes of $10^6$ to $10^9$ solar masses,
as it is believed exist at the center of
galaxies such as M87, M51 and others \cite{Ginzburg}, present much
better conditions to delay the gravitational waves relatively to the 
electromagnetic waves. After all, the Schwarzschild radius of a $10^9$
solar mass black hole has the size of two Astronomical Units,
a distance that light takes about 1000 seconds to cross if traveling
outside the horizon. This may give us enough time for the potential to
act on the gravitational waves, slowing them down for sufficient
time, in order to accumulate a time delay possible to be measured
with the technology of the next generations of gravitational wave 
observatories and electromagnetic telescopes in the few decades to come.

\section{Acknowledgements}

We would like to thank the participants of the ``Pequeno Semin\'ario'' of 
Lafex/CBPF. This work was supported by ``Conselho Nacional de Desenvolvimento 
Cient\'{\i}fico e Tecnol\'ogico'' (CNPq) and ``Funda\c{c}\~ao de 
Amparo a Pesquisa no Estado do Rio de Janeiro'' (FAPERJ) of Brazil.

\end{document}